\documentclass[runningheads]{llncs}

\usepackage[T1]{fontenc}
\usepackage{todonotes}
\usepackage{amsmath}
\usepackage{amsfonts}
\usepackage{xspace}
\usepackage{booktabs}
\usepackage{tabularx}
\usepackage{colortbl}
\usepackage{caption}
\usepackage{subcaption}
\usepackage{algorithm}
\usepackage[noend]{algpseudocode}
\PassOptionsToPackage{hyphens}{url}\usepackage{hyperref}
\usepackage{cleveref}
\usepackage{subcaption}
\usepackage{nicefrac}
\usepackage{tikz}
\usepackage{multicol,multirow}
\usepackage{pgfplots}
\usepackage{pifont}
\usepackage{microtype}
\usepackage{csquotes}
\usepackage{siunitx}
\pgfplotsset{compat=1.16}
\usepgfplotslibrary{statistics}
\usetikzlibrary{positioning,arrows.meta,arrows,shapes.misc,intersections,fit,decorations.text}
\usetikzlibrary{decorations.pathreplacing,calligraphy,spy}

\def\prot{\textsc{Pirates}\xspace}
\def\gaprot{GAddra\xspace}
\def\gmkg{\textit{GMK}_g\xspace}
\def\pir{\textsc{PIR}\xspace}
\def\co{\textit{C}\bar{O}_{C^\textit{SR}}\xspace}

\newtheorem{assumption}{Assumption}

\definecolor{color1}{rgb}{145,167,221}
\definecolor{color2}{HTML}{a8b7df}
\definecolor{color3}{HTML}{bee3e3}
\definecolor{color4}{HTML}{d3ecca}
\definecolor{color5}{HTML}{f5f7d4}

\newcommand{\furl}[1]{\footnote{\url{#1} --- Accessed April 13, 2024}}

\begin{document}

\title{\prot: Anonymous Group Calls Over Fully Untrusted Infrastructure}
\subtitle{Extended Version}


\author{
    Christoph Coijanovic\inst{1}\orcidID{0000-0002-5873-2859} \and
    Akim Stark\inst{2} \and
    Daniel Schadt\inst{1}\orcidID{0009-0009-6357-1314} \and
    Thorsten Strufe\inst{1}\orcidID{0000-0002-8723-9692}
}

\authorrunning{C. Coijanovic et al.}

\institute{
    Karlsruhe Institute of Technology (KIT), Karlsruhe, Germany \email{firstname.lastname@kit.edu} \and
    FZI Forschungszentrum Informatik, Karlsruhe, Germany \email {lastname@fzi.de}
}

\maketitle

\begin{abstract}
Anonymous metadata-private voice call protocols suffer from high delays and so far cannot provide group call functionality.
Anonymization inherently yields delay penalties, and scaling signalling and communication to groups of users exacerbates this situation.
Our protocol \prot employs PIR, improves parallelization and signalling, and is the first group voice call protocol that guarantees the strong anonymity notion of communication unobservability.
Implementing and measuring a prototype, we show that \prot with a single server can support group calls with three group members from an 11 concurrent users with mouth-to-ear latency below 365ms, meeting minimum ITU requirements as the first anonymous voice call system.
Increasing the number of servers enables bigger group sizes and more participants.
\end{abstract}


\vspace{0.5cm}
\textit{%
    This is the extended version of ``\textsc{Pirates}: Anonymous Group Calls Over Fully Untrusted Infrastructure'', published at ACISP 2024.
    Compared to the conference version, this version contains an evaluation of worker scalability (\Cref{sec:eva:scale}) and dialing performance (\Cref{sec:eva:dialing}), as well as a discussion on fixed versus dynamic groups (\Cref{sec:discussion}). 
}
\vspace{0.5cm}

\section{Introduction}
\label{sec:intro}
The Coronavirus pandemic has significantly changed the way people communicate with each other.
Work-from-home and quarantine regulations prevented in-person meetings, thus people shifted their communication online:
Video conferencing service Zoom saw a 30 times increase in daily users just in the first few months of the pandemic\footnote{\url{https://www.npr.org/2021/03/19/978393310/}, accessed \today}.
However, Zoom and similar services do not have a great track record of privacy and security~\cite{RefZoomSec1,RefZoomSec2,RefZoomSec3}.

While some mainstream services today offer end-to-end encryption for confidentiality, they still disclose \emph{metadata} (e.g., who calls whom).
This threatens privacy, as metadata can reveal most sensitive information, e.g., health conditions, sexual orientation, or political views~\cite{RefPhonePrivacy}.
Recently, a first generation of anonymous voice communication protocols emerged, namely Addra~\cite{RefAddra} and Yodel~\cite{RefYodel}.
However, no anonymous protocol so far supports group calls, a central feature of mainstream services.
In this paper, we propose \prot, the first protocol to provide anonymity for group calls over a fully untrusted infrastructure.

Most existing anonymous communication networks require \emph{some} trust in intermediary systems;
For example, the \emph{Anytrust} model~\cite{RefVuvuzela,RefD3,RefTalek} requires that at least one arbitrary server providing the service remains honest.
In light of powerful adversaries like nation states that cooperate with ISP and cloud providers this seems a strong assumption, especially knowing, that they are explicitly targeting anonymous communication\footnote{\url{https://www.theguardian.com/world/2013/oct/04/tor-attacks-nsa-users-online-anonymity}, accessed \today}. \stepcounter{footnote}
\footnotetext{\url{https://www.itu.int/rec/T-REC-G.114-200305-I/}, accessed \today}
\addtocounter{footnote}{-1}
Thus, \prot forgoes any trust assumptions about its infrastructure and can guarantee anonymity even if all servers and intermediaries are malicious.

\prot, like most related protocols, operates in rounds.
Each round, clients send short voice snippets to their mailbox at a central server.
From there, the other clients in that call can retrieve the mailbox content anonymously using private information retrieval (\pir)~\cite{RefCPIR}.
The voice snippets from all call participants are combined and played to the user.

In providing anonymous group calls (multicast), \prot faces multiple challenges:
First, voice communication requires low latency.
ITU recommendation G.114 states that \emph{mouth-to-ear} latency for telephony applications should not exceed 400 ms.\footnotemark
Mouth-to-ear latency is defined as the time between a word being spoken at one end and and it being heard a the other.
This measure of latency includes the time needed to record, transmit and playback.
Addra and Yodel only evaluate transmission latency, which is not enough to determine suitability for voice communication.
When considering mouth-to-ear latency, neither meets the recommendation with their evaluated parameters.
Second, the overhead for multicast is much higher compared to unicast calls, as data from every call participant has to be distributed to every other call participant.
Third, more metadata (e.g., the number of participants in a call) can leak to the adversary.

In our design of \prot, we combine a number of concepts to improve privacy and performance. 
Following previous approaches, we choose PIR as the fundamental anonymization primitive. 
Each client sends her voice data to their own mailbox, from which other clients on the same call can retrieve it anonymously. 
\pir allows clients to hide from which mailboxes they retrieve packets, even if the server storing the mailbox is malicious.

To reduce the overhead that is inherent to group calls, we follow Angel et al.'s idea of multi-retrieval \pir~\cite{RefPung}.
The server divides a single, large database into multiple smaller {\em buckets}, from which clients retrieve in parallel.
This reduces both computation and communication, as buckets naturally contain fewer entries.
To implement \pir, we chose a FastPIR~\cite{RefAddra} after performing an empirical evaluation of the most promising contestants from recent related work.

With respect to privacy, we aim to hide \emph{all} metadata including if any real communication occurs.
To do so, we require clients to send cover traffic to appear in a call even if they are not.
Fixing the maximum group size to \(G\) ensures that the overhead remains acceptable.

To finally minimize the amount of data that needs to be transmitted between clients, \prot uses the LPCNet vocoder~\cite{RefLPC}, which requires only 1.6 kbit/s per client for speech transmissions.

To summarize, our main contributions in this paper are:
\begin{itemize}
    \item The design and implementation of \prot, the first group call protocol that offers strong metadata protection over fully untrusted infrastructure.
    \item Formal proof that \prot achieves communication unobservability~\cite{RefNotions}.
    \item In-depth experimental evaluation of \prot, showing that it achieves sub-400-ms mouth-to-ear latency.
\end{itemize}

\section{Related Work}
\label{sec:related}
There are a plethora of anonymous communication protocols.
However, most achieve one-way latency in the order of seconds (Vuvuzela~\cite{RefVuvuzela}, Stadium~\cite{RefStadium}, Express~\cite{RefExpress}, Pung~\cite{RefPung}, Karaoke~\cite{RefKaraoke}, 2PPS~\cite{Ref2PPS}, Loopix~\cite{RefLoopix}), minutes (XDR~\cite{RefXDR}, Blinder~\cite{RefBlinder}, Clarion~\cite{RefClarion}, Atom~\cite{RefAtom}), or even hours (Riposte~\cite{RefRiposte}, Spectrum~\cite{RefSpectrum}) for large numbers of users and are therefore not usable for real-time communication.

The anonymous communication service Tor~\cite{RefTor} offers latency suitable for voice communication~\cite{RefShorTor,RefDonar} and can be extended to support multicast~\cite{RefMTor}, but is vulnerable to attacks from malicious entry/exit routers, onion routers, and even observation on network links~\cite{RefTorSurvey}.
Therefore, Tor cannot provide the strong privacy protection that we require. 
Several protocols promise anonymous voice communication protected against stronger adversaries than Tor's, which we will discuss in greater detail.

Drac~\cite{RefDrac} lets users communicate through onion-encrypted~\cite{RefOnion} messages over a friend-to-friend network.
However, as it requires trust in the relays, it is not suitable for the settings we consider. 
Further, the authors give no empirical data validating the suitability for voice communication.

Herd~\cite{RefHerd}, Yodel~\cite{RefYodel}, and Hydra~\cite{RefHydra} are based on mix networks~\cite{RefMixnet}.
In a mix network, servers called \emph{mixes} batch and shuffle user requests, add noise by creating fake requests and route the messages forward, while removing layers of encryption.
These mixes can then be chained together, and ensure unlinkability between sender and receiver, if at least one (arbitrary) mix between them is honest.
Indeed, Herd requires trust in the first server of the chain, Yodel assumes that each server only has a 20\% chance of being corrupted, while Hydra vaguely assumes `a limited number of system entities' to be malicious.
As we aim to enable anonymous communication over \emph{fully} untrusted infrastructure, mix-based approaches are insufficient for our purpose.

Frank and Sorger~\cite{RefFS} aim to achieve anonymous voice communication based on DC-nets~\cite{RefDCNet}.
Compared to mix networks, DC-nets require no trusted servers, as messages are sent from peer to peer.
However, a certain fraction of peers has to be honest.
To hide the link between two communication partners, three other peers (besides the communication partners themselves) have to be honest in their case.
DC-nets additionally suffer from poor scalability.

Addra~\cite{RefAddra} is the closest relative to our work:
Clients retrieve messages using computational private information retrieval (CPIR)~\cite{RefCPIR}.
CPIR allows Addra to make strong privacy guarantees, even if the infrastructure and all users except the communication partners are malicious.
Addra only supports one-to-one communication natively.

Finally, all related work only evaluate transmission latency rather than mouth-to-ear latency.
Without an evaluation of mouth-to-ear latency, their actual suitability for voice communication is not clear.

\section{Private Information Retrieval}
\label{sec:background}
We want to give a brief overview of private information retrieval (\pir), as it forms the basis of \prot.
\pir schemes come in two flavors:
Information-theoretic~\cite{RefITPIR1,RefITPIR2,RefITPIR3} and computational~\cite{RefCPIR,RefSpiral,RefAddra,RefCPIR2,RefOPIR}. 
Computational \pir schemes are secure under some cryptographic hardness assumption (e.g., LWE~\cite{RefLWE}).
Information-theoretic schemes achieve that the best attack even of an adversary with infinite resources is random guessing.
We focus on computational \pir, as information-theoretic schemes require multiple servers of which at least one must be honest.
This requirement does not align with our threat model.

\pir schemes assume a database of items stored on a remote server.
Clients want to access specific items without revealing to the server which items they are interested in.
If \pir is used for communication, where receivers fetch the content written by a sender, this mechanism effectively leads to the \emph{unlinkability} of sender and receiver.
A \pir scheme consists of five algorithms (adapted from~\cite[Def. 2.3]{RefSpiral}):
\begin{itemize}
    \item \textsc{Setup}$(1^\lambda, 1^N) \to (pk, sk)$.
        On input of the security parameter $\lambda$ and a bound on the database size \(N\), \textsc{Setup} returns a key pair $(pk, sk)$.
    \item \textsc{Send}$(m,i)$.
        On input of a message $m$ and index $i\in\{1,\dots,N\}$, the $i$th database item $d_i$ is replaced by $m$.
    \item \textsc{Query}$(sk,i) \to (st, q)$.
        On input of a secret key $sk$ and an index $i\in\{1,\dots,N\}$, \textsc{Query} outputs a state $st$ and a query $q$.
    \item \textsc{Answer}$(pk, db, q) \to r$.
        On input of a public key $pk$, a database $db = \{d_1,\dots,d_N\}$, and a query $q$, \textsc{Answer} outputs a response $r$. 
    \item \textsc{Decode}$(sk, st, r) \to d_j$.
        On input of a secret key $sk$, a state $st$ and a response $r$, \textsc{Decode} outputs a database item $d_j \in db$. 
\end{itemize}
If the \pir scheme is \emph{correct}, \textsc{Decode} will return the database item matching the index the client has requested (i.e., $j = i$) assuming all input is valid.
If the \pir scheme is \emph{private}, the adversary cannot gain any information about the contained index from receiving and processing a query $q$.

When \pir is used for communication, clients need a way to update database items (to send a message).
These updates are done by sending data, together with the index it should be written to, directly to the server.
As \pir stores the database in plain, clients have to encrypt their data to ensure confidentiality.

One can build a na\"ive computational \pir scheme using \emph{fully homomorphic} encryption (FHE)~\cite{RefHE}.
\textsc{Query}$(sk, i)$ returns a vector $\vec{v} = \{v_1,\dots,v_N\}$, where each component is a FHE ciphertext.
The $i^{th}$ index is an encryption of `1', whereas all other entries are encryptions of `0'\footnote{
    \pir requires IND-CPA-secure encryption to ensure that 0- and 1-entries are indistinguishable.}.
The server generates a vector $\vec{r} = \{r_1,\dots,r_N\}$ by interpreting the plaintext database items as integers and computing $r_j \gets v_j \cdot d_j$ for all $j\in\{1,\dots,N\}$.
Then it computes $r \gets \sum_{j=1}^N r_j$ and sends it to the client.
Because of the homomorphic properties, the client can decrypt $r$ to reveal the desired database item.

\section{Model \& Goals}
\label{sec:goals}
\subsection{Setting}
\label{sec:goal:sec}
We assume that participants of \prot are organized in groups with at most $G$ members.
We assume that all members of a group $g$ know each other (and their respective public keys) and share a symmetric secret group master key $\gmkg$.

Further, we make two assumptions about the used cryptographic building blocks:
\begin{assumption}
    \label{ass:hash}
    We assume that $H$ is a cryptographic hash function with preimage resistance.    
\end{assumption}
Preimage resistance requires that the adversary, given a hash value $H(x)$, cannot invert the function to retrieve its input, or, in other words, does not learn information about the preimage $x$.
Preimage resistance is a standard property of cryptographic hash functions such as SHA-3~\cite{RefSHA}.

\begin{assumption}
    \label{ass:enc}
    We assume clients have access to an IND-CPA secure symmetric encryption scheme.
\end{assumption}
If an encryption scheme achieves IND-CPA security, ciphertexts do not reveal any information about the contained plaintexts.
AES in CBC mode fulfills this requirement~\cite{RefAES}.

Finally, we assume that the clients' and servers' clocks are \emph{somewhat} synchronous.
Current research suggests that is is possible to synchronize clocks over the internet to within 1 ms.\furl{https://engineering.fb.com/2020/03/18/production-engineering/ntp-service/}
This is sufficient for our usecase.
Note that out-of-sync clients are not able to communicate with others, but do not endanger the privacy of themselves or others.

\subsection{Threat Model}
\label{sec:goal:threat}
\prot aims to protect the privacy of honest clients against a \emph{global} and \emph{active} adversary $\mathcal{A}$ who may be in control of \emph{all} infrastructure (i.e., \prot's servers and all infrastructure used to connect to them).
The adversary may further control an arbitrary number of clients.
Note that we only guarantee privacy for groups of trusted and benign users, which do not contain compromised members that are under control of the adversary.
The adversary can simply leak group membership and decrypt voice messages due to common system functionality, otherwise. 
The adversary’s active abilities allow it to drop, delay, insert, and modify any packet. 

\subsection{Privacy Goal}
\label{sec:goal:pricay}
We do not want $\mathcal{A}$ to learn \emph{any} information about the communication activities between \emph{honest} clients.
We can formalize this privacy goal using Kuhn et al.'s privacy notion of \emph{Communication Unobservability} ($\co$)~\cite{RefNotions}.
Kuhn et al. defined communication unobservability based on a game played between the adversary $\mathcal{A}$ and a challenger $\mathcal{C}$.
$\mathcal{A}$'s task is to distinguish between two self-chosen \emph{scenarios} (each consisting of a series of communications) based on the protocol's output which it receives from $\mathcal{C}$.
The subscript $C^\textit{SR}$ denotes that $\mathcal{A}$ may not corrupt clients who are part of the scenarios' communications.
$\mathcal{A}$ may corrupt other protocol participants arbitrarily.
A protocol achieves $\co$ if there is no efficient $\mathcal{A}$ who can win the associated game with a non-negligible advantage over random guessing.

Kuhn et al. have designed their privacy notions for one-to-one communication.
Thus, every communication is defined by a single sender, receiver, and message.
We will map \prot's group calls to this communication format by breaking them down to their equivalent unicasts:
If $\mathcal{A}$ wants Alice, Bob, and Carol to be in a group call in one scenario, the scenario would contain communications between each pair of them (i.e., Alice$\to$Bob, Alice$\to$Carol, Bob$\to$Alice, Bob$\to$Carol, Carol$\to$Alice, and Carol$\to$Bob).
For each resulting unicast communication, the message contains the sender's voice data of that round.

\subsection{Non-Goals}
\label{sec:goal:non}
Like related works~\cite{RefYodel,RefAddra}, \prot aims to hide communication patterns rather than protocol participation.
To ensure that the mere usage of \prot does not raise suspicion, should be usable for everyday-communication;
Our goal is to provide call quality comparable to mainstream solutions.

In this paper, we focus on communication within fixed and preexisting groups.
We consider the bootstrapping and management of groups an interesting but orthogonal problem.
Na\"ively, groups can be set up via in-person meetings.
The adaption of anonymous bootstrapping services such as Alpenhorn~\cite{RefAlpenhorn} is also possible.

We assume a powerful adversary who is able to mount denial of service (DoS) attacks.
In case of such an attack, we do not aim to ensure availability.
\prot's privacy guarantees, on the other hand, must still hold.

\section{System Design}
\label{sec:arch}
In this section, we give a high-level overview of \prot by describing its goals and how they are achieved.

\subsection{Group Call Functionality}
\prot aims to enable calls within groups of clients.
A client can be part of multiple groups, but only be in an active call with one group at a time.
Thus, two main protocol functions are required:
1) Clients need to be able to inform other group members of their intend to call.
We call this process \emph{dialing}.
If a client receives call requests from more than one group, she has to decide which call to participate in.
2) Once a call is ongoing, voice data from every group member has to be transmitted to every other group member.

In \prot, dialing and communication are split into separate \emph{phases}.
First, all clients participate in a dialing phase, after which the actual calls are executed.
The communication phase is further divided into short rounds, as data is not sent as a continuous stream but rather in small packets;
Clients send one packet per communication round.
The number of communication rounds is fixed through a protocol parameter.
After a communication phase ended, a new dialing phase starts.
We call the combination of a dialing phase and its subsequent communication phase an \emph{epoch}.
Calls automatically end with the communication phase, but can be picked up again in the next epoch.

\subsection{Hiding Metadata}
In both dialing and communication, \prot aims to hide \emph{all} metadata, even if all infrastructure is controlled by malicious entities.

\prot hides metadata during dialing with a novel invite mechanism:
A client $A$ who wants to start a call in group $g$ generates an \emph{invite} that is deterministically derived from 1) a secret shared by all group members, 2) $A$'s identity and 3) the current epoch.
The shared group secret is included so that only group members can recognize the invite, it appears random to any other party.
The client's identity is included to hide the number of invitations a group receives in a given dialing phase.
The current epoch is included in the invite to unlink invites from the same client to a group over multiple dialing phases.
Clients who do not want to start a call generate a random, cover invite of equal length.
Thus, every client generates exactly one invite per dialing phase which is then sent to the server.
The server broadcasts the collected invites to all clients, who process them locally to determine if they have an incoming call.

To hide communication metadata, \prot uses a system of mailboxes.
Every communication round, each clients sends one fixed-size encrypted voice packet \emph{to their own} mailbox.
If a client is not on a call, they send a dummy voice packet to their mailbox.
Similarly, if a group wants to end their call before the end of the communication phase, its members have to send dummy voice packets for the remaining rounds.
This approach ensures that the clients' sending behavior is independent of their actual communication patterns.

On the receiving side, every client in a call needs to fetch the mailbox contents from each other group member.
\prot hides who is fetching which mailboxes through the use of private information retrieval (\pir).
Clients can submit a query for each mailbox, the answering server only learns that a query was made, not which mailbox it targets.
Clients also always submit as many queries as the maximum group size.
If they are in a call with fewer members or no call at all, queries to random mailboxes are generated.

\subsection{Improving Performance}
\pir answers are expensive to compute.
Scalability of \prot would be severely limited if a single server was responsible for all queries.

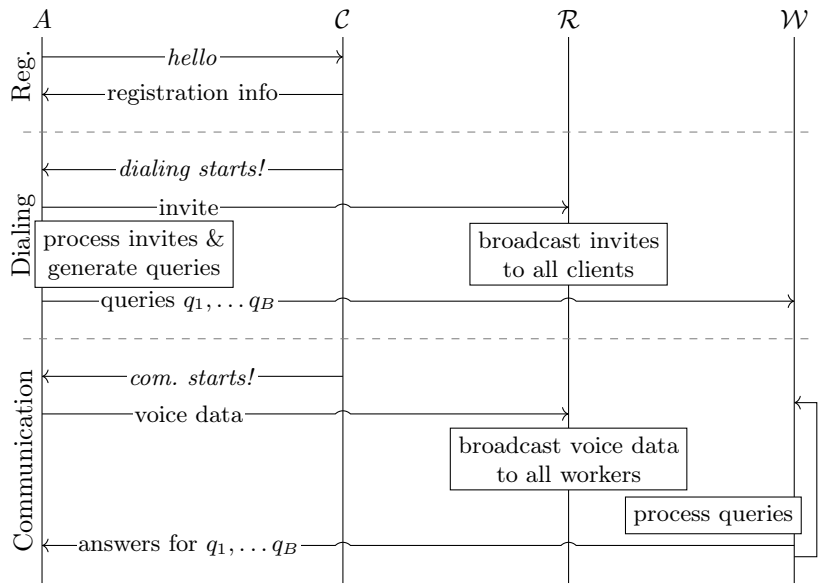
\begin{figure}
    \centering
    \begin{tikzpicture}[font=\footnotesize]
    \node[font=\normalsize] (a) at (0,0) {$A$};
    \node[font=\normalsize] (c) at (4,0) {$\mathcal{C}$};
    \node[font=\normalsize] (r) at (7,0) {$\mathcal{R}$};
    \node[font=\normalsize] (w) at (10,0) {$\mathcal{W}$};
    
    \draw[] (a) --++(0,-7.5);
    \draw[] (c) --++(0,-7.5);
    \draw[] (r) --++(0,-7.5);
    \draw[] (w) --++(0,-7.5);
    
    \draw[->] (0,-0.5) -- node[midway,fill=white,inner sep=0.5] {\textit{hello}} (4,-0.5);
    \draw[->] (4,-1) -- node[midway,fill=white,align=center,inner sep=0.5] {registration info} (0,-1);
    \node[rotate=90,font=\normalsize] (l1) at (-0.25,-0.75) {Reg.};
    
    \draw[->] (4,-2) -- node[midway,fill=white,inner sep=0.5] {\textit{dialing starts!}} (0,-2);
    \draw[->] (0,-2.5) -- node[midway,fill=white,inner sep=0.5] {invite} (3.9,-2.5) to[out=45,in=135] (4.1,-2.5) -- (7,-2.5);
    \node[draw,fill=white,align=center] (a1) at (7,-3.125) {broadcast invites\\to all clients};
    \node[draw,fill=white,align=center,anchor=west] (a2) at (-0.1,-3.125) {process invites \&\\generate queries};
    \draw[->] (0,-3.75) -- node[midway,fill=white,inner sep=0.5] {queries $q_1,\dots q_B$} (3.9,-3.75) to[out=45,in=135] (4.1,-3.75) -- (6.9,-3.75) to[out=45,in=135] (7.1,-3.75) -- (10,-3.75);
    \node[rotate=90,font=\normalsize] (l2) at (-0.25,-2.875) {Dialing};
    
    \draw[->] (4,-4.75) -- node[midway,fill=white,inner sep=0.5] {\textit{com. starts!}} (0,-4.75);
    \draw[->] (0,-5.25) -- node[midway,fill=white,inner sep=0.5] {voice data} (3.9,-5.25) to[out=45,in=135] (4.1,-5.25) -- (7,-5.25);
    \node[draw,fill=white,align=center] (a3) at (7,-5.85){broadcast voice data\\to all workers};
    \node[draw,fill=white,align=center,anchor=east] (a4) at (10.1,-6.6){process queries};
    \draw[<-] (0,-7) -- node[midway,fill=white,inner sep=0.5] {answers for $q_1,\dots q_B$} (3.9,-7) to[out=45,in=135] (4.1,-7) -- (6.9,-7) to[out=45,in=135] (7.1,-7) -- (10,-7);
    \draw[->] (10,-7.15) --++(0.3,0) --++(0,2.05) --++(-0.3,0);
    \node[rotate=90,font=\normalsize] (l3) at (-0.25,-5.875) {Communication};
    
    \draw[dashed,black!50] (-0.25,-1.5) -- (10.25,-1.5);
    \draw[dashed,black!50] (-0.25,-4.25) -- (10.25,-4.25);
\end{tikzpicture}
    \caption{
        Simplified interaction between a client \(A\), and \prot's servers (coordinator \(\mathcal{C}\), relay \(\mathcal{R}\), and worker \(\mathcal{W}\)) 
    }
    \label{fig:arch}
\end{figure}

Thus, \prot divides the single \emph{logical} server into a hierarchy of multiple \emph{physical} servers.
At the top of the hierarchy sits a single \emph{coordinator}.
The coordinator manages communication by announcing the start of each phase.
When a client first joins the system, she registers at the coordinator, which assigns her a \emph{relay} and a \emph{worker} server.
Each relay server is responsible for handling messages from a fixed subset of clients.
In the dialing phase, clients send invites to their relay, which broadcasts them globally.
After processing the invites, clients generate \pir queries and send them to their worker.
In the communication phase, clients send voice packets to their relay.
The relay forwards its clients' packets to all \emph{workers}, who compute replies based on the data and the previously received queries.
The replies are sent back to the expecting clients.
\Cref{fig:arch} presents an overview of the tasks of the coordinator, relays, and workers.

As we have discussed above, each client needs to receiver the data of every other group member ($G-1$ in total).
\prot implements a multi-query \pir~\cite{RefPung}, to reduce the cost of each query.
The database items are distributed evenly into $B$ \emph{buckets}.
Angel et al. suggest $B$ to equal $1.5$ times the number of retrievals per client.
Instead of querying the entire database $G-1$ times, clients submit one query per bucket.
As each bucket contains significantly fewer items than the entire database computing queries and answers is cheaper.
In their evaluation, Angel et al. determine that for one million database items and 16 retrievals per client, latency decreases by a factor of $1.5\times$ compared to the na\"ive approach, even if answers are computed in series.
As the buckets are disjoint, a client's answers can also be processed perfectly in parallel, reducing the response time for all of a client's queries to that of a single one (given enough resources).

If each item is put into exactly one bucket, situations can occur where a client cannot retrieve all items she is interested in.
As the client can only retrieve one item per bucket, two items of interest in the same bucket cannot be retrieved simultaneously.
The likelihood of such a collision can be reduced by putting each item into multiple buckets.
That way, clients have a chance to retrieve one of the items in case of a collision from another bucket.

\section{Protocol Specification}
\label{sec:protocol}
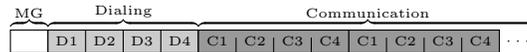
\begin{figure}
    \centering
\scalebox{1}{
\begin{tikzpicture}
    \draw[black,fill=black!0] (0,0) rectangle (0.5,0.3);
    \draw[black,fill=black!20] (0.5,0) rectangle (2.5,0.3);
    \newcounter{ga}\setcounter{ga}{1};
    \foreach \i in {1,1.5,2,2.5}{
        \draw[black] (\i,0) -- (\i,0.3);
        \node[font=\tiny] at (\i-0.25,0.15) {{D\thega}};
        \stepcounter{ga};
    }
    \draw[black,fill=black!40] (2.5,0) rectangle (4.5,0.3);
    \draw[] (3,0) -- (3,0.2);
    \draw[] (3.5,0) -- (3.5,0.2);
    \draw[] (4,0) -- (4,0.2);
    \node[font=\tiny] at (2.75,0.15) {C1};
    \node[font=\tiny] at (3.25,0.15) {C2};
    \node[font=\tiny] at (3.75,0.15) {C3};
    \node[font=\tiny] at (4.25,0.15) {C4};
    
    \draw[black,fill=black!40] (4.5,0) rectangle (6.5,0.3);
    \draw[] (5,0) -- (5,0.2);
    \draw[] (5.5,0) -- (5.5,0.2);
    \draw[] (6,0) -- (6,0.2);
    \node[font=\tiny] at (4.75,0.15) {C1};
    \node[font=\tiny] at (5.25,0.15) {C2};
    \node[font=\tiny] at (5.75,0.15) {C3};
    \node[font=\tiny] at (6.25,0.15) {C4};
    
    \node[font=\tiny] at (6.75,0.15) {\(\cdots\)};
    \draw[thick,decorate,decoration = {calligraphic brace}] (0,0.35) -- node[midway,above] {\tiny MG} (0.5,0.35);
    \draw[thick,decorate,decoration = {calligraphic brace}] (0.5,0.35) -- node[midway,above] {\tiny Dialing} (2.5,0.35);
    \draw[thick,decorate,decoration = {calligraphic brace}] (2.5,0.35) -- node[midway,above] {\tiny Communication} (7,0.35);
\end{tikzpicture}
}
    \caption{
        \prot operates in epochs that are split into three phases:
        Mapping Generation, Dialing, and Communication.
        The communication phase consists of multiple subrounds.
        Not to scale, size of phase does not correspond to duration.
    }
    \label{fig:phases}
\end{figure}

Having introduced the design of \prot in §\ref{sec:arch}, we now describe the protocol in more detail.
\prot operates in \emph{epochs}.
Each epoch is split into three phases: 
Mapping Generation, Dialing, and Communication.
The communication phase is further divided into multiple short rounds where one voice packet is exchanged per round.
\Cref{fig:phases} gives a visual overview of an epoch in \prot.

\subsection{Registration}
\label{sec:prot:reg}
Before a client can participate in the protocol, she has to register at the coordinator $\mathcal{C}$. 
Upon sending a `hello', the client is assigned a relay $\mathcal{R}_i$ and a worker server $\mathcal{W}_i$ and informed of the assignment.
The client further receives a mailbox identifier, an authentication token, and the total number of mailboxes $N$.
$\mathcal{C}$ informs $\mathcal{R}_i$ and $\mathcal{W}_i$ of the new client assigned to them.
$\mathcal{R}_i$ receives $A$'s mailbox identifier and authentication token.
During registration, \prot's servers learn who is \emph{participating} in the protocol.
As related work, \prot does not aim to hide participation, but rather communication patterns between participants.

\subsection{Mapping Generation}
\label{sec:prot:mg}
During the Mapping Generation phase, the coordinator $\mathcal{C}$ distributes $N$ mailboxes into $B$ buckets.
To select buckets for each mailbox, we use a variant of 3-way Cuckoo hashing~\cite{RefCuckoo}: 
\begin{enumerate}
    \item $\mathcal{C}$ generates a random seed $s$ which is used to initialize three hash functions $h_0, h_1, h_2$.
    Each hash function takes as input a mailbox identifier $i\in\{1,\dots,N\}$ and a nonce $n\in\mathbb{Z}$.
    It outputs a bucket $b\in\{1,\dots,B\}$.
    \item For each mailbox $i$, $\mathcal{C}$ computes $b_0 \gets h_0(i \mid n), b_1 \gets h_1(i \mid n)$, and $b_2 \gets h_2(i \mid n)$ to derive three distinct buckets the mailbox is mapped to.
    The nonce $n$ is initially set to 0 and incremented if the hash functions do not produce distinct buckets.
    $\mathcal{C}$ sends the ordered list of mailboxes in each bucket to every worker.
    \item $\mathcal{C}$ publishes the seed $s$.
\end{enumerate}

\subsection{Dialing}
\label{sec:prot:dial}
The dialing phase is split into four subphases D1 to D4.
D1 to D3 cover the invitation process, whereas \pir queries are generated and distributed in D4.

\paragraph{D1 -- Invite Sending}
Each client $A$ generates an invite $h$.
If $A$ wants to initiate a call in group $g$, she computes $h \gets H(\gmkg \mid pk_A \mid e)$, where $H(\cdot)$ is a cryptographic hash function, $\gmkg$ is the group's shared secret, $pk_A$ is her public key, and $e$ is the current epoch's number.
If $A$ does not want to initiate a call, she computes $h \gets H(r)$, where $r$ is a random string of equal length to $\gmkg \mid pk_A \mid e$.
The client $A$ then sends her invite to her relay $\mathcal{R}_i$.

As we also need a random initialization vector (IV) for the use of the block cipher, we interpret the resulting hash $h$ as such.
If multiple invites for the same group are received, users choose the hash with the lowest interpreted binary value as IV.

\paragraph{D2 -- Invite Broadcast}
After a fixed amount of time, each relay assembles all received invites into a package that is broadcast to all participants (globally, not just to the clients assigned to that relay).

\paragraph{D3 -- Invite Processing}
Let $I = \{i_1,\dots,i_N\}$ be the set of invitations the clients received from the relays.
Assume group $g$ has members $A$, $B$, and $C$.
$C$ wants to check if there is an invite for group $g$:
\begin{enumerate}
    \item $C$ computes the reference invites $i_{\textit{ref},A} \gets H(\gmkg\mid pk_\mathbf{A}\mid e)$ and $i_{\textit{ref}, B} \gets H(\gmkg\mid pk_\mathbf{B}\mid e)$
    \item $C$ checks if $i_{\textit{ref},A} \in I$ or $i_{\textit{ref},B}\in I$.
        If either check succeeds, $C$ expects a call in group $g$.
\end{enumerate}
If $C$ expects call in multiple groups, she has to decide which single dialing request to accept.
For groups whose dialing request is not accepted, no further action is needed.
For the group whose dialing request is accepted, $C$ has to proceed to subphase D4.

\paragraph{D4 -- Query Generation}
Assume that $C$ accepts a dialing request in group $g$ with other members $A$ and $B$.
$C$ has to generate \pir queries for the mailboxes of $A$ and $B$:
\begin{enumerate}
    \item $C$ initializes the hash functions $h_0, h_1, h_2$ with seed $s$
    \item $C$ generates the mapping from mailbox to buckets locally for all mailboxes
    \item $C$ selects an index $i$ for each bucket $b$.
        $C$ tries to find a combination of indices so that she can retrieve the mailboxes of $A$ and $B$ (and random mailboxes from the remaining buckets).
        If this should not be possible, $C$ cannot join the call and selects random indices for all buckets instead.
    \item $C$ generates a \pir query for each bucket requesting the corresponding index
    \item $C$ sends her queries to $\mathcal{W}_i$, the worker assigned to her
\end{enumerate}
If $C$ does not expect a call, she behaves analogously but selects a random index for every bucket by default.

\subsection{Communication}
\label{sec:prot:com}
Each communication round is split into phases C1 to C4.

\paragraph{C1 -- Snippet Sending}
Let $m$ be the client $A$'s voice data gathered since the last C1 phase for a call in group $g$.
\begin{enumerate}
    \item $A$ computes the ciphertext $c\gets \text{Enc}(\gmkg, m)$
    \item $A$ sends $c$ along with her mailbox ID and auth token to her relay $\mathcal{R}_i$.
    \item $\mathcal{R}_i$ deposits $c$ into $A$'s mailbox if the provided authentication token matches the one given out previously.
\end{enumerate}

\paragraph{C2 -- Distribute Ciphertexts}
After waiting a fixed amount of time, each relay broadcasts the mailbox contents (along with the corresponding mailbox identifiers) to all workers.
For mailboxes where the relay expected a ciphertext but did not receive one, the relay sends random data.

\paragraph{C3 -- Answering Queries}
Based on the received mailbox data and the mailbox-to-bucket mapping, each worker assembles the buckets.
Then, the workers compute \pir answers to the queries assigned to it.
The workers send the answers to the expecting clients.

\paragraph{C4 -- Decode}
The client $A$ discards any answers from queries to random indices.
For all other answers, $A$ applies the \pir decode procedure and decrypts the recovered ciphertext using $\gmkg$.
The decrypted voice snippets from her communication partners are overlayed and output.

\section{Privacy Analysis}
\label{sec:proof}
We want to provide formal proof that \prot achieves strong privacy protection against the assumed adversary $\mathcal{A}$ (see \Cref{sec:goal:threat}).

\prot's privacy depends on that of the underlying \pir scheme.
Intuitively, \pir's privacy goal requires that a malicious server gains no information about the index a given query is requesting to retrieve.
This can be formalized as a game between a challenger and the adversary, where the adversary submits two challenge indices $\textit{idx}_0, \textit{idx}_1$ and receives a query for a randomly chosen $\textit{idx}_b$ for $b\in\{0,1\}$.
The adversary's task is then to determine the value of $b$.
For reference, we present Menon et al.'s formal definition of query privacy~\cite[Def. 2.3]{RefSpiral}:

\begin{definition}[Query Privacy]
    \label{def:qp}
    For all polynomials $N = N(\lambda)$ and all efficient adversaries $\mathcal{A}$, there exists a negligible function $\text{negl}(\cdot)$ such that for all $\lambda\in \mathrm{N}$, \(\left| Pr\left[ \mathcal{A}^{\mathcal{O}_b(qk,\cdot,\cdot)}(1^\lambda,pp) = b\right] -\frac{1}{2}\right| = \text{negl}(\lambda)\) where $(pp,qk) \gets \textsc{Setup}(1^\lambda,1^N), b\gets^R \{0,1\}$, and the oracle $\mathcal{O}_b$ $(qk,\textit{idx}_0,\textit{idx}_1)$ outputs $\textsc{Query}(qk,\textit{idx}_b)$.
\end{definition}

\begin{lemma}
    \label{lem:proof:pir}
    FastPIR achieves query privacy.
\end{lemma}
Ahmad et al. argue that FastPIR achieves query privacy by reducing its security to that of the underlying BFV encryption scheme~\cite{RefAddra}.

With that, we can now prove \prot's privacy protection.
We do so via a series of hybrid games.
The first hybrid is equivalent to the $\co$ game played with standard \prot.
Each hybrid differs from the previous in that it provides less information that depends on the executed scenario to the adversary.
In the final game, all output the adversary receives will be independent from the scenario selected.
Thus $\mathcal{A}$ can only guess at random in the final hybrid and not have an advantage in winning the $\co$ game.
For each pair of subsequent games, we show that there is no distinguisher $\mathcal{D}$ who can --- based on the output of $\mathcal{A}$ --- distinguish which hybrid is executed with an advantage over random guessing.
If we show the impossibility of such a $\mathcal{D}$ for every step between hybrids, we have also shown that $\mathcal{A}$ can only resort to random guessing in the first hybrid, which is equivalent to the $\co$ game with \prot.

For better readability, we extract the proof of indistinguishability for each pair of hybrids into \Cref{lem:privacy:01,lem:privacy:12,lem:privacy:23,lem:privacy:34,lem:privacy:45}.
The rest of the proof is presented in \Cref{thm:privacy}.

\begin{theorem}
    \label{thm:privacy}
    \prot achieves $\co$ against $\mathcal{A}$.
\end{theorem}
\begin{proof}
Let $H_0,\dots,H_5$ be the following series of hybrid games:
\begin{itemize}
    \item $H_0$ --- \textit{Original game}:
        $\co$ with \prot as specified in \Cref{sec:protocol}.
    \item $H_1$ --- \textit{No registration}:
        As $H_0$, but $\mathcal{A}$ receives no output during registration.
    \item $H_2$ --- \textit{No mapping generation}:
        As $H_1$, but $\mathcal{A}$ receives no output during the mapping generation phase.
    \item $H_3$ --- \textit{Random invites}:
        As $H_2$, but all clients compute cover invites (i.e., $h\gets H(r)$ where $r$ is a random string of fixed length).
    \item $H_4$ --- \textit{Random queries}:
        As $H_3$, but all clients compute \pir queries to random indices.
    \item $H_5$ --- \textit{Random messages}:
        As $H_4$, but all clients send random messages.
\end{itemize}
We prove that in hybrid $H_5$, any observation the adversary can make is \emph{independent} from the communications specified by the challenge's chosen scenario:
By definition, $H_5$ does have no observable output during registration and mapping generation, so we do not need to consider these steps here.

In subphase D1, all participants (including challenge clients) behave identically:
Each participant sends a hashed random string of fixed length to their designated relay.

In subphase D2, all invites are broadcast to all participants.
As the number of participants is independent of the adversary's challenge and each participant sent exactly one invite in the previous subphase, protocol behavior in this subphase does not depend on the challenge scenario.

Subphase D3 occurs completely within the clients.
As $\co$ does not allow $\mathcal{A}$ to corrupt any challenge client, $\mathcal{A}$ receives no output from this subphase.

In subphase D4, every participants generates and outputs exactly one \pir query to random index for each bucket, independent of specified communication patterns.

In phase C1, every participant sends a fixed-size ciphertext of random data to their own mailbox their relay, independent of the specified communication patterns.

In phase C2, the relays distribute the ciphertexts to the workers.
As the number of participants is independent of the adversary's challenge and each participant sent exactly one ciphertext in the previous phase, protocol behavior in this phase does not depend on the challenge scenario.

In phase C3, the workers compute \pir queries and send the resulting answers to the participants.
As each participant receives has sent the same amount of queries to random indices in subphase D4, the workers computations do not depend on the challenge scenario.

Finally, phase C4 occurs completely within the clients. 
As $\co$ does not allow $\mathcal{A}$ to corrupt any challenge client, $\mathcal{A}$ receives no output from this phase.

We have shown that there cannot be any efficient $\mathcal{A}$ who can distinguish between scenarios in hybrid $H_5$ with a non-negligible advantage over random guessing.
\Cref{lem:privacy:01,lem:privacy:12,lem:privacy:23,lem:privacy:34,lem:privacy:45} iteratively prove that in this case, there also cannot be an efficient $\mathcal{A}$ who can distinguish between scenarios in hybrid $H_0$ with a non-negligible advantage over random guessing. 
As $H_0$ is identical to the $\co$ game with full \prot, we have proven the theorem.
\end{proof}

\begin{lemma}
    \label{lem:privacy:01}
    There is no efficient distinguisher $\mathcal{D}$ who can distinguish $\mathcal{A}$'s output in hybrid $H_0$ from $\mathcal{A}$'s output in $H_1$ with a non-negligible advantage over random guessing.
\end{lemma}
\begin{proof}
    $H_1$ only differs from $H_0$ in that $\mathcal{A}$ receives no output during registration in $H_0$.
    Recall that registration has to be executed once by every protocol participant \emph{prior} to communication.
    As the protocol participants have to be identical in both scenarios, registration does not change depending on the scenario.
    Thus, any advantage that $\mathcal{A}$ has in guessing the correct scenario in $H_0$, it still has in $H_1$, so no $\mathcal{D}$ can distinguish $\mathcal{A}$'s output depending on the hybrid.
\end{proof}

\begin{lemma}
    \label{lem:privacy:12}
    There is no efficient distinguisher $\mathcal{D}$ who can distinguish $\mathcal{A}$'s output in hybrid $H_1$ from $\mathcal{A}$'s output in $H_2$ with a non-negligible advantage over random guessing.
\end{lemma}
\begin{proof}
    $H_2$ differs from $H_1$ in that $\mathcal{A}$ receives no output during the mapping generation phase in $H_2$.
    We can argue analogously to the proof of \Cref{lem:privacy:01}, as mapping generation is independent of communication patterns.
    Thus, there cannot be a distinguisher $\mathcal{D}$ which can distinguish between $\mathcal{A}$'s output in $H_1$ versus $H_2$.
\end{proof}
\begin{lemma}
    \label{lem:privacy:23}
    There is no efficient distinguisher $\mathcal{D}$ who can distinguish $\mathcal{A}$'s output in hybrid $H_2$ from $\mathcal{A}$'s output in $H_3$ with a non-negligible advantage over random guessing.
\end{lemma}
\begin{proof}
    Assume that there is such a distinguisher $\mathcal{D}$.
    This implies that there is an adversary $\mathcal{A}$, which can produce a non-negligible difference in the  outputs of game $H_2$ and $H_3$.
    The \emph{only} difference between these games is the structure of the invites:
    While in $H_2$ invites of challenge clients are hashes of their key material plus current epoch number, they are hashes of random strings.
    To gain an advantage from what is output by the hash function, $\mathcal{A}$ would have to break \emph{preimage resistance}, which contradicts \Cref{ass:hash}.
    Since there cannot be such an $\mathcal{A}$, there also cannot exist $\mathcal{D}$.
\end{proof}
\begin{lemma}
    \label{lem:privacy:34}
    There is no efficient distinguisher $\mathcal{D}$ who can distinguish $\mathcal{A}$'s output in hybrid $H_3$ from $\mathcal{A}$'s output in $H_4$ with a non-negligible advantage over random guessing.
\end{lemma}
\begin{proof}
    Assume that there is such a distinguisher $\mathcal{D}$.
    This implies that there is an adversary $\mathcal{A}$, which can produce a non-negligible difference in the  outputs of game $H_3$ and $H_4$.
    The \emph{only} difference between these games is in the indices of the \pir requests of challenge clients:
    While in $H_3$ the \pir requests are to the indices of the other group members' mailboxes, they are to random indices in $H_4$.
    To gain an advantage from the index contained in the \pir queries, $\mathcal{A}$ would have to break the \pir scheme's \emph{query privacy}.
    This contradicts however Ahmad et al.'s proof of query privacy for FastPIR (which is used by \prot)~\cite[Sec. 4.4]{RefAddra}.
    Since there cannot be such an $\mathcal{A}$, there also cannot exist $\mathcal{D}$.
\end{proof}
\begin{lemma}
    \label{lem:privacy:45}
    There is no efficient distinguisher $\mathcal{D}$ who can distinguish $\mathcal{A}$'s output in hybrid $H_4$ from $\mathcal{A}$'s output in $H_5$ with a non-negligible advantage over random guessing.
\end{lemma}
\begin{proof}
    Assume that there is such a distinguisher $\mathcal{D}$.
    This implies that there is an adversary $\mathcal{A}$, which can produce a non-negligible difference in the  outputs of game $H_4$ and $H_5$.
    The \emph{only} difference between these games is in content of ciphertext from challenge clients:
    While in $H_4$ clients encrypt and send the messages specified in the challenge communications, they generate random ones in $H_5$.
    To gain an advantage from the ciphertexts' content, $\mathcal{A}$ would have to break the encryption scheme's IND-CPA security, which contradicts \Cref{ass:enc}.
    Since there cannot be such an $\mathcal{A}$, there also cannot exist $\mathcal{D}$.
\end{proof}

\section{Evaluation}
\label{sec:eva}
In this section we investigate the performance of \prot.
Due to space constraints, we focus the evaluation on our main hypothesis:
\emph{\prot enables group calls with a mouth-to-ear latency of less than 400 ms.}
    
To evaluate this hypothesis, we must first determine which \pir scheme is best suited for use with \prot.
We do so in \Cref{sec:eva:pir}.
We further evaluate \prot's scalability in the number of workers in \Cref{sec:eva:scale} and the overhead of \prot's dialing protocol in \Cref{sec:eva:dialing}.

\subsection{\pir Schemes}
\label{sec:eva:pir}
We first evaluate the performance of the state-of-the-art \pir schemes to determine which is best suited for \prot.
We consider FastPir~\cite{RefAddra}, Spiral~\cite{RefSpiral} (stream variant), and OnionPIR~\cite{RefOPIR} using their prototype implementations run on a laptop with an AMD Ryzen 5 5625U.
To put the performance of these computational \pir schemes into perspective, we also include information-theoretic \pir~\cite{RefITPIR1} and a broadcast of the entire database in the comparison.
For each scheme, we use a database of 64 elements, each 96 bytes in size, as this corresponds to our exptected database size for \prot{}.
We compare the time to preprocess the database, compute a response, send that response to the callee, and decode the response, as these are the steps that affect latency in \prot.
We limit the callee's bandwidth to 20 Mbit/s to model typical mobile device bandwidth.

\newcolumntype{R}{>{\raggedleft\arraybackslash}X}
\begin{table}[]
    \centering
    \begin{tabularx}{\textwidth}{lRRRRR}\toprule
    & \cite{RefAddra} & \cite{RefSpiral} & \cite{RefOPIR} & BC & IT-PIR\\\midrule
    Prepro. (ms) & 1.913 & 70.035 & 0.000 & 0.000 & 0.0000\\
    Reply (ms) & 10.689 & 15.418 & 792.000 & 0.000 & 0.0387 \\
    Send (ms) & 25.600 & 4.400 & --- & 2.143 & 0.0004\\
    Decode (ms) & 0.448 & 0.265 & --- & 0.000 & 0.0002\\\addlinespace
    \textbf{Total (ms)} & 38.650 & 90.118 & 792.000 & 2.143 & 0.0393\\\bottomrule
    \end{tabularx}
    \caption{
        Comparison of \pir schemes.
        OnionPIR's prototype does not output reply size or decode time, hence the `---' in the respective columns.
    }
    \label{tab:pir_comparison}
\end{table}

\Cref{tab:pir_comparison} shows our results.
We see that IT-PIR achieves by far the lowest total latency because it is computationally cheap and produces a small response.
However, IT-PIR's trust model does not match that of \prot{}, as it requires multiple servers, at least one of which is honest.
For the database size we consider, broadcasting the entire database has lower latency than computational PIR, since it requires no processing.
However, broadcasting does not scale well for larger databases and imposes high bandwidth costs on clients. 
Of the three computational PIR schemes considered, FastPIR~\cite{RefAddra} has the lowest total latency and is therefore used for our \prot{} prototype.

\subsection{Mouth-to-Ear Latency}
\label{sec:eva:latency}
To be suitable for voice communication, a system must achieve low \emph{mouth-to-ear} latency (i.e., the time between a person speaking and the other person hearing their voice).
The ITU recommends a latency of less than 400 ms.
We evaluate for which parameters \prot meets this recommendation.

We have written a prototype implementation of \prot in C++, which can be found on GitHub\footnote{\url{https://github.com/kit-ps/pirates}}.
For all our measurements, we use a server with an AMD Ryzen 9 7950X3D and 128GB RAM.
For homomorphic encryption, we use Microsoft's SEAL library v4.1.1.
LPCNet (FreeDV fork) is used as the speech encoder.

For all experiments, we used networked Docker containers to simulate multiple machines.
We ran one container as caller, one as relay, one as worker, and one as callee.
We measured the latency between when the caller starts encoding her voice snippet and when the callee finishes decoding the snippets of all group members.
To derive the mouth-to-ear latency, we added the following additional delays to our measurements
\begin{enumerate}
    \item \emph{One snippet length} to account for recording and playback of the snippet.
    \item \emph{15ms} to account for network latency, which is not present in our dockerized setup.
        Based on Verizon roundtrip latency measurements within Europe\furl{https://www.verizon.com/business/terms/latency/}, we assume 7ms one-way latency between caller and relay and between worker and callee.
        We also assume 1ms delay between relay and worker to represent the situation found in current data centers.
    \item \emph{10ms} to account for the audio processing latency of modern mobile devices\furl{https://superpowered.com/superpowered-android-media-server}.
\end{enumerate}
To simulate additional users of the system, the relay duplicates the caller's snippet into a bucket of size \(\left\lceil 3\cdot\frac{\textit{\# Users}}{\textit{\# Buckets}}\right\rceil\) where \(\textit{\# Buckets} = \lceil 1.5\cdot(G-1)\rceil\).
The worker will process \textit{\# Buckets} \pir requests for each client, where the callee is set to a random index between 0 and \textit{\# Users}.
The \pir requests are distributed among 12 threads, while additional available threads are used to parallelize the response computation.

We are investigating the effects of group size and number of clients on mouth-to-ear latency.
For each combination of parameters, we need to find the optimal snippet length:
Short snippets reduce the overall latency because the caller has less to buffer, but the worker has less time to process.
If the worker cannot finish processing in time for the next snippet, the audio playback will be interrupted.
A quick poll of colleagues found that interruptions up to \SI{10}{\ms} are acceptable (see \Cref{apx:survey}), giving the worker some leeway.

We run two experiments:
In the first experiment, we vary the group size between 2 and 5 with a fixed number of clients of 6.
In the second experiment, we vary the number of clients between 3 and 12 with a fixed group size of 3. 
For each combination of parameters, we test snippet lengths between 40ms and 300ms in steps of 20ms to determine the optimal value (i.e., the shortest snippet with a ratio between mean worker processing time and snippet length of at most 1.1).
We repeat each measurement 90 times (plus 10 warm-up rounds) and present the mean and standard deviation.

We expect latency to increase linearly with group size, since group size determines the number of buckets, which determines the number of \pir requests.
We expect latency to increase super-linearly with the number of users, since more users both increase the number of \pir requests and make each \pir response more expensive to compute due to the increased bucket size.

\begin{figure}[t]
    \begin{subfigure}[c]{0.5\textwidth}
    \centering
    \begin{tabularx}{\textwidth}{XXXXX}\toprule
         \rotatebox{90}{Group Size} & \rotatebox{90}{\textbf{M2E (ms)}} & \rotatebox{90}{Std. Dev.} & \rotatebox{90}{Snip. (ms)} &\rotatebox{90}{Ratio}\\\midrule
         2 & 204.19 & 16.39 & 60 & 0.86\\
         3 & 214.51 & 24.55 & 60 & 1.02\\
         4 & 275.89 & 39.47 & 80 & 1.09\\
         5 & 569.89 & 99.08 & 200 & 1.06\\\bottomrule
    \end{tabularx}
    \vspace{1cm}
    \caption{
        Fixed number of clients of 6.
    }
    \label{tab:group_size_latency}
    \end{subfigure}
    \begin{subfigure}[c]{0.5\textwidth}
    \begin{tikzpicture}[thick, scale=0.7]
\begin{axis}[
  xlabel={Number of Clients},
  x label style={at={(axis description cs:1.0,0.02)},anchor=south east},
  ylabel={Mouth-to-Ear Latency (ms)},
  legend pos=north west,
  legend style={draw=none}]
\addplot+[
  mark options={black, scale=1},
  mark=o,
  only marks,
  error bars/.cd, 
    y fixed,
    y dir=both, 
    y explicit
] table [x=x, y=y,y error=error, col sep=comma] {
    x, y, error
    3, 205.378, 15.13
    4, 205.90, 14.86
    5, 211.92, 18.78
    6, 214.51, 24.55
    7, 256.8, 28.8
    8, 259.18, 31.54
    9, 263.39, 35.39
    10, 317.17, 42.07
    11, 364.36, 57.31
    12, 413.86, 66.22
    13, 459.28, 92.16
};
\addplot [] coordinates {(3.3,205)} node[rotate=90,font=\footnotesize]{60ms, 0.87};
\addplot [] coordinates {(4.3,205)} node[rotate=90,font=\footnotesize]{60ms, 0.88};
\addplot [] coordinates {(5.3,211)} node[rotate=90,font=\footnotesize]{60ms, 0.97};
\addplot [] coordinates {(6.3,214)} node[rotate=90,font=\footnotesize]{60ms, 1.02};
\addplot [] coordinates {(7.3,256)} node[rotate=90,font=\footnotesize]{80ms, 0.96};
\addplot [] coordinates {(8.3,259)} node[rotate=90,font=\footnotesize]{80ms, 0.98};
\addplot [] coordinates {(9.3,263)} node[rotate=90,font=\footnotesize]{80ms, 1.04};
\addplot [] coordinates {(10.3,317)} node[rotate=90,font=\footnotesize]{100ms, 1.1};
\addplot [] coordinates {(11.3,364)} node[rotate=90,font=\footnotesize]{120ms, 1.08};
\addplot [] coordinates {(12.3,413)} node[rotate=90,font=\footnotesize]{140ms, 1.09};
\addplot [] coordinates {(13.3,459)} node[rotate=90,font=\footnotesize]{160ms, 1.07};
\end{axis}
\end{tikzpicture}
    \caption{
        Fixed group size of 3.
        Labels denote snippet length and ratio.
    }
    \label{fig:com:clients}
    \end{subfigure}
    \caption{
        Mouth-to-ear latency measurements in \prot for varying numbers of clients (a) and varying group size (b).
        Ratio refers to the ratio between worker processing time and snippet length.
    }
\end{figure}
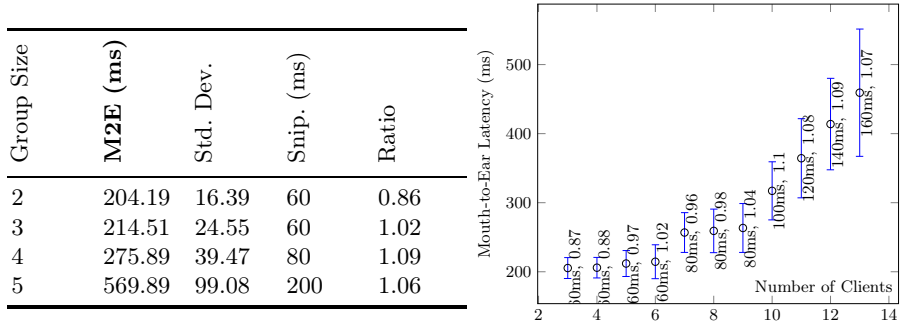

\Cref{tab:group_size_latency} shows our measurements of the impact of group size on mouth-to-ear latency.
As the group size increases, the worker's computation time also increases due to the additional queries for the new group members.
The longer the worker needs to compute, the longer the snippets have to be, which further increases the mouth-to-ear latency directly (due to the recording/playback delay) and indirectly (due to the larger size of the longer snippets).
This behavior can be seen in the distinct jumps in mouth-to-ear latency as the snippet length increases, especially as the group size increases from 4 to 5.

\Cref{fig:com:clients} shows our measurements of the effect of the number of clients on mouth-to-ear latency.
As expected, mouth-to-ear latency grows superlinearly with the number of clients in the system.
Additional clients increase the worker's processing time, which requires longer snippets.
We also note that the standard deviation grows with the number of clients.
This is due to the larger range from which the callee's position in the worker queue is selected.

To better understand the causes for \prot's mouth-to-ear latency, we break down the latency of three parameter combination into the following protocol steps:
LPCNet encoding, symmetric encryption, transmission from caller to relay, transmission from relay to worker, database preprocessing, \pir response computation, transmission to callee, \pir decoding, symmetric decryption, and LPCNet decoding.
We select one parameter combination with long snippets (denoted LS -- 200ms snippets, groups of 3, 6 clients), one with large groups (denoted LG -- 80ms snippets, groups of 5, 6 clients), and one with a large number of clients (denoted MC -- 80ms snippets, groups of 3, 11 clients).
With all combinations, we expect \pir response computation to most expensive protocol step.

\Cref{tab:com_eval} presents the broken down mouth-to-ear latency.
Considering only the \emph{transmission} latency, our expectations are confirmed that \pir response computation has by far the largest impact on latency, followed by \pir decoding.
However, especially in the LS scenario, the recording and playback of the snippet itself outweighs even the impact of \pir (see the `\textit{Additional}' column).
Note that our choice of parameters results in interrupted playback for the LG and MC scenarios.

\begin{table}[]
    \centering
    \begin{tabularx}{\textwidth}{lXXXXXXXXXXXr}\toprule
    & \rotatebox{90}{Vo. Enc.} & \rotatebox{90}{Encrypt} & \rotatebox{90}{C\(\to\)R} & \rotatebox{90}{R\(\to\)W} & \rotatebox{90}{Prepro.} & \rotatebox{90}{\pir Rep.} & \rotatebox{90}{W\(\to\)C} & \rotatebox{90}{\pir Dec.} & \rotatebox{90}{Decrypt} & \rotatebox{90}{Vo. Dec} & \rotatebox{90}{\textit{Add.}} &  \rotatebox{90}{\textbf{Total}}\\
    \midrule
    \textbf{LS} & 14.14 & 0.04 & 11.72 & 1.56 & 13.68 & 81.53 & 1.5 & 47.13 & 0.03 & 37.52 & 225 & \textbf{433.84}\\
    \textbf{LG} & 5.57 & 0.02 & 1.21 & 11.32 & 6.31 & 107.02 & 1.82 & 47.01 & 0.04 & 29.7 & 105 & \textbf{315.01}\\
    \textbf{MC} & 5.97 & 0.02 & 0.96 & 11.14 & 6.73 & 91.55 & 1.42 & 46.83 & 0.03 & 17.23 & 105 & \textbf{286.98}\\
    \bottomrule
    \end{tabularx}
    \caption{
        Contribution of individual protocol steps to mouth-to-ear latency in ms for three different sets of parameters.
        The `\textit{Add.}' step combines network latency, audio stack latency, and recording/playback time.
    }
    \label{tab:com_eval}
\end{table}

In summary, our evaluation of mouth-to-ear latency confirms that \prot is the first protocol suitable for anonymous group calling.
A single worker can support 6 clients in groups of 4\footnote{
   Since each client can only be in one active call at a time, there can only be one group in call among the 6 clients. 
   The remaining 2 clients will still behave as if they were in an active call with 3 other members.}
or up to 11 clients in groups of 3 with less than 400ms mouth-to-ear latency.

\subsection{Worker Scalability}
\label{sec:eva:scale}
Ahmad et al. found that using more worker servers in Addra only reduces latency for up to 80 workers~\cite[Sec. 6.1]{RefAddra}.
For more workers, the master (which has to distribute the client's data to all workers) becomes a bottleneck and latency increases again.
\prot aims to overcome this bottleneck by splitting the data distribution between several relay servers.
Our goal in this section is to evaluate whether the relays actually allow \prot to benefit from a larger number of workers compared to Addra.
For a sensible comparison, we assume a group size of two (i.e., one-to-one communication) for \prot in the following section.

Note that the total time taken to send and process voice snippets is made up of four components:
1) The time it takes the clients to send their snippet to the master/relay,
2) The time it takes the master/relay to send its collected snippets to all workers,
3) the time it takes the worker to compute all the answers based on the data, and
4) the time it takes the worker to send the answers back to the clients.
We can approximate the client send time by dividing the size of the snippet by the client's bandwidth, and similarly the worker send time of the answers by dividing the total size of all its computed answers by the worker's bandwidth.
Similarly, the send time of the master/relay can be approximated by dividing the total amount of data it has to send by its bandwidth.
The amount of data each relay has to send depends on the number of workers, clients and relays, as well as the size of each snippet:
\[
    \#\text{Workers}\cdot\left(\frac{\#\text{Clients}\cdot\text{Snippet}}{\#\text{Relays}}\right)   
\]
For Addra, the number of relays is one.
The computation time per worker can be calculated by multiplying the number of clients by the time it takes to compute a response (since each client is waiting for a response) and dividing the result by the number of workers, which is the number of requests each worker can answer in parallel.

We assume that the snippets are 400 bit (250 ms snippet length, 1.6 Kb/s data rate for LPCNet) and that the clients are connected to the servers via a 100 Mb/s connection.
Based on our measurements in \Cref{sec:eva:pir}, we assume that it takes a worker 13 ms to compute a \pir response that is 64 KB in size.
Based on the setup used to evaluate Addra, we assume that each worker can compute 48 answers in parallel and that the servers have 12 Gb/s connections between them.

To compare Addra's approach with \prot's, we set the number of clients in both cases to $2^{15}$, vary the number of workers between 20 and 200, and compute the sending/computing time as described above.
For \prot we use one relay per 20 workers.

For Addra, we expect to observe a similar infliction point as Ahmad et al., while for \prot, we expect a negative correlation between the number of workers and the sending/computing time.

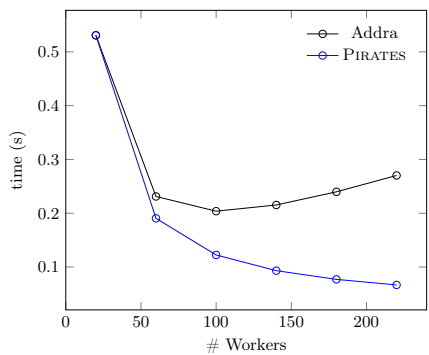
\begin{figure}
    \centering
    \begin{tikzpicture}[
spy using outlines={
    rectangle,
    magnification=3,
    connect spies,
},
thick,
scale=0.7,
]
\begin{axis}[
  xlabel={\# Workers},
  ylabel={time (s)},
  legend pos=north east,
  legend style={draw=none}]
\addplot+[
  black, 
  mark options={black, scale=1},
  mark=o,
] table [x=x, y=y, col sep=comma] {
    x, y
    20, 0.530748866780599
    60, 0.23117230428059896
    100, 0.2038090751139323
    140, 0.21533346499488468
    180, 0.23982039455837675
    220, 0.27019938761393225
};
\addlegendentry{Addra}
\addplot+[
  blue, 
  mark options={blue, scale=1},
  mark=o,
] table [x=x, y=y, col sep=comma] {
    x, y
    20, 0.530748866780599
    60, 0.1904822001139323
    100, 0.12242886678059896
    140, 0.09326315249488466
    180, 0.07705997789171007
    220, 0.06674886678059896
};
\addlegendentry{\prot}

\end{axis}
\end{tikzpicture}
    \caption{Correlation between the number of workers and the total time it takes to distribute voice data and compute \pir answers in \prot and Addra.}
    \label{fig:eva:workers}
\end{figure}
\Cref{fig:eva:workers} presents our results.
As we expected, \prot's total time to distribute the data and compute the answer decreases continuously as the number of workers increases, while Addra's time initially decreases but then increases again.
Note that our results show a higher inflection point for Addra (100 workers) compared to Ahmed et al.'s evaluation (80 workers).
This discrepancy can be explained by our assumed snippet length of 250 ms.
Our calculation also shows an infliction point of around 80 workers if we assume the same 480 ms snippets as Ahmed et al., which validates our method.

Based on our results, we can strongly confirm our hypothesis:
\prot's latency can be reduced by introducing more workers without restriction.
In fact, further analysis shows that latency continues to decrease until there are as many workers as clients.

\subsection{Dialing}
\label{sec:eva:dialing}
We suggest an improvement of the Dialing protocol in \Cref{sec:prot:dial}, which we investigate next.
The time it takes clients to create invites is 1) independent of other protocol variables (as every client has to create exactly one invite per dialing phase) and 2) negligible (as only a single small ciphertext/hash has to be computed).
We hence focus on the time it takes clients to process the received invites and compare the invite mechanism of \prot to a modified Addra mechanism:
In Addra, the invites consist of `hello' messages encrypted with the communication partner's public key.
A na\"ive adaption to group calls would require computing separate invites for every group member.
As a simple improvement, we encrypt the `hello' message with $\gmkg$ in \gaprot instead, so that there is only one possible invite per group which other group members need to check for.

We first investigate how group size impacts the time it takes clients to process invites.
We set the number of protocol participants to $2^{15}$ and vary the group size between $2$ and $2048$.
Such big groups are probably unrealistic for anonymous voice communication.
We still want to analyze the behaviour with growing number of invites.

For \prot, one would expect processing time to scale linearly in the group size.
As each client computes an individual invite (by including their public key into the hash), group members need to compare each received invite to $G-1$ possible invites.
With data structures that support lookups in constant time, the processing is very efficient.
Computing the intersection between the received and expected invites has an average time complexity in $\mathcal{O}(G-1)$.
For \gaprot, the processing time is independent of the group size.
This is due to the fact that there is only one possible invite per group: $Enc_{\gmkg}(\textit{hello})$.

The results in \Cref{fig:eval:dial_members} indeed show that
 \prot's processing is orders of magnitude faster than \gaprot's:
Even for 2048 group members, \prot takes only 28.5 $\mu$s, dropping to 0.677$\mu$s with $G=8$, where \gaprot takes a constant 71 ms.

In our second experiment, we want to determine the impact the number of protocol participants has on the processing time of clients.
For both \prot and \gaprot, we fix the group size at 4 and vary the number of protocol participants between $2^{12}$ and $2^{22}$.
\Cref{fig:eval:dial_active} presents our results.
In both \prot's and \gaprot's mechanism, the processing time scales linearly in the number of participants.
This is to be expected, as there is one invite to be checked for every participant and invites can be checked independently from each other.
However, \prot's processing takes significantly less time:
For $2^{16}$ participants, \prot requires 0.6 $\mu$s, whereas \gaprot requires 135 ms.
This significant difference can be explained by processing that each entry requires:
In \gaprot, each invite has to be decrypted, whereas, in \prot, clients only need to compute the intersection between received invites and expected ones.
\begin{figure}
    \begin{subfigure}[c]{0.5\textwidth}
        \begin{tikzpicture}[thick, scale=0.7]
\begin{axis}[
  xtick={2,4,8,16,32,64,128,256,512,1024,2048}, 
  xticklabels={$2^{1}$, $2^{2}$, $2^{3}$, $2^{4}$, $2^{5}$, $2^{6}$, $2^{7}$, $2^{8}$, $2^{9}$, $2^{10}$, $2^{11}$},
  xmode=log,
  xlabel={\# group members},
  ylabel={time ($\mu$s)},
  legend pos=north west,
  legend style={draw=none}]
\addplot+[
  black, 
  mark options={black, scale=1},
  mark=o,
  smooth, 
  error bars/.cd, 
    y fixed,
    y dir=both, 
    y explicit
] table [x=x, y=y,y error=error, col sep=comma] {
    x, y, error
    2, 6.340199999932406e-01, 5.815492714689167e-01
4, 6.381599999971677e-01, 5.277736682010419e-01
8, 6.768400000012331e-01, 5.07675724070909e-01
16, 7.143699999911934e-01, 5.898692999404881e-01
32, 8.943600000044239e-01, 7.617806826627724e-1
64, 1.0757900000035737e-0, 7.844335187551419e-01
128, 1.8145099999999913e-0, 1.5237766047991177e-0
256, 2.5429200000026685e-0, 1.4280164577533033e-0
512, 4.988360000003356e-0, 2.732168992652086e-0
1024, 9.239880000000422e-0, 4.823249921335239e-0
2048, 28.505860000005433e-0, 8.228762939562392e-0
};
\end{axis}
\end{tikzpicture}
        \caption{Invite processing time for $2^{15}$ protocol participants and group sizes between $2$ and $2048$ (time in $\mu$s).}
        \label{fig:eval:dial_members}
    \end{subfigure}
    \begin{subfigure}[c]{0.5\textwidth}
        \begin{tikzpicture}[thick, scale=0.7]
\begin{axis}[
  xtick={4096,8192,16384,32768,65536,131072,262144,524288,1048576,2097152,4194304}, 
  xticklabels={$2^{12}$, $2^{13}$, $2^{14}$, $2^{15}$, $2^{16}$, $2^{17}$, $2^{18}$, $2^{19}$, $2^{20}$, $2^{21}$,$2^{22}$},
  xmode=log,
  xlabel={\# participants},
  ylabel={time (s)},
  legend pos=north west,
  legend style={draw=none}]
\addplot+[
  black, 
  mark options={black, scale=1},
  mark=o,
  smooth, 
  error bars/.cd, 
    y fixed,
    y dir=both, 
    y explicit
] table [x=x, y=y,y error=error, col sep=comma] {
    x, y, error
4096, 6.157899999996441e-07, 5.155159089916009e-07
8192, 7.218899999986151e-07, 1.0696842628203785e-06
16384, 6.208999999979259e-07, 5.796974045177533e-07
32768, 6.195599999991863e-07, 5.26421698300961e-07
65536, 6.304699999937213e-07, 4.3594218073319166e-07
131072, 6.564800000052884e-07, 6.171117227838369e-07
262144, 6.477900000145808e-07, 5.428789193609328e-07
524288, 7.235500000923878e-07, 1.2242521971055301e-06
1048576, 6.310600000958289e-07, 6.340518383170586e-07
2097152, 6.55750000149169e-07, 7.077845275061483e-07
4194304, 7.064800007583471e-07, 9.257799928533341e-07
};
\addlegendentry{\prot}
\addplot+[
  blue, 
  mark options={blue, scale=1},
  mark=triangle,
  smooth, 
  error bars/.cd, 
    y fixed,
    y dir=both, 
    y explicit
] table [x=x, y=y,y error=error, col sep=comma] {
    x, y, error
4096, 0.008377547260000002, 6.983269662031317e-05
8192, 0.017212040389999995, 4.768595155510843e-05
16384, 0.034042469790000016, 0.0007851456851116794
32768, 0.06935331294000001, 0.0004418580685990355
65536, 0.13510565172999994, 0.0028126794168162315
131072, 0.2762964803100001, 0.004819313119115407
262144, 0.5750629344599999, 0.00963831604918373
524288, 1.1413954010999996, 0.0074504507620431785
1048576, 2.4085921841299998, 0.030926233631072424
2097152, 4.760146183540002, 0.040160009796065044
4194304, 8.610711484171429, 0.07562980246758891
};
\addlegendentry{Addra/\gaprot}
\end{axis}
\end{tikzpicture}
        \caption{Invite processing time for groups of 8 members and $2^{12}$ to $2^{17}$ protocol participants (time in seconds).}
        \label{fig:eval:dial_active}
    \end{subfigure}
\end{figure}

\section{Fixed vs. Dynamic Groups}
\label{sec:discussion}
As discussed in \Cref{sec:goal:sec}, \prot requires that clients share a symmetric secret and know each others' public keys before they can start a group call.
This complicates the situation in settings that require spontaneous group formation.
One could for example imagine an anonymous crisis hotline where people in need of immediate physical or psychological help can call their choice of medical professionals.
To adapt \prot for such settings the dialing phase has to be modified so that group members can derive a shared secret just in time.
We cannot use our hash-based approach for this purpose, as clients no longer know which invites to expect.
To avoid the need for prior knowledge of public keys, identity-based encryption (IBE)~\cite{RefIBE} can be used instead:
With IBE, a sender can derive a public key for an intended receiver from any identifier (e.g., phone number, email address).

Assume client $A$ wants to start a call with $B$ and $C$.
$A$ first derives $pk_B$ and $pk_C$ from $B$'s and $C$'s identities respectively.
She assembles an invite that includes a `hello' message as well as the public keys of all group members (including her own).
$A$ then encrypts the invite once with $pk_B$ and once with $pk_C$ and sends the two resulting ciphertexts to the server.
The server broadcasts the invites to all clients, who try to decrypt each one with their secret key.
$B$ and $C$ will succeed in decrypting $A$'s invite.
They can use the included public keys and their secret key to derive a shared $\gmkg$.

As we have shown in \Cref{sec:eva:dialing}, Addra's encryption-based approach to dialing requires significant overhead.
Thus, for \prot, we focus on settings with static groups, where lightweight dialing can be used.

\section{Conclusion}
\label{sec:conclusion}
We have introduced \prot, the first anonymous communication network supporting group calls.
We provided formal proof that \prot discloses no information about the communication patterns between honest users, even if the infrastructure between them is in full control of the adversary.
Through empirical evaluation, we have shown that \prot reaches acceptable mouth-to-ear latency, through database sharding and probabilistic batch codes.
While \prot introduces significant overhead compared to non-anonymous solutions, it proves that anonymous \pir-based group calls are possible with today's technology.
Currently, \prot might not be suitable for some settings, e.g., calls between geographically distant clients with high network latency between them.
As we have shown, most of \prot's overhead compared to non-anonymous solution stems from its \pir scheme.
If future \pir schemes are more efficient, \prot will be able to use them, thus increasing its efficiency in turn.

\subsection*{Acknoledgements}
This work has been funded by the Helmholtz Association through the KASTEL Security Research Labs (HGF Topic 46.23), and by funding of the German Research Foundation (DFG, Deutsche Forschungsgemeinschaft) as part of Germany's Excellence Strategy -- EXC 2050/1 -- Project ID 390696704 -- Cluster of Excellence \enquote{Centre for Tactile Internet with Human-in-the-Loop} (CeTI) of Technische Universität Dresden.

\bibliographystyle{splncs04}
\bibliography{ref}

\appendix

\section{Audio Quality Survey}
\label{apx:survey}
If user in \prot does not receive the next voice snippet before the current snippet finished playback, the audio stream is interrupted.
We want to determine how much interruption, if any, are users willing to tolerate before being dissatisfied with the audio quality.

We recorded a 89s voice sample which we split into 100ms snippets.
We then created five versions of the audio file, which differed in the amount of silence that was inserted between snippets.
One file had 2ms inserted, one 5ms, one 10ms, one 20ms, and one 30ms.
We presented 19 users with the files and asked them which file they would consider to have the minimum acceptable quality for an audio call.

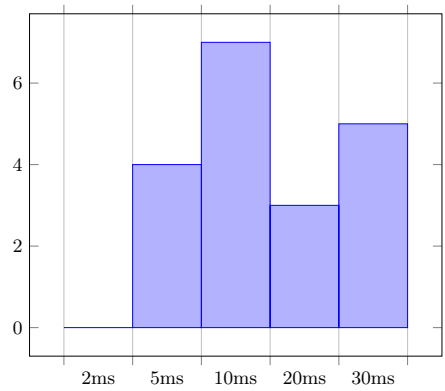
\begin{figure}
    \centering
    \begin{tikzpicture}[scale=0.8]
        \begin{axis}[
            ybar interval, 
            xticklabels={2ms, 5ms, 10ms, 20ms, 30ms},
            ]
            \addplot coordinates { (0, 0) (1, 4) (2, 7) (3, 3) (4, 5) (5, 0)};
        \end{axis}
        \end{tikzpicture}
    \caption{
        Results of interrupted playback survey.
    }
    \label{fig:survey}
\end{figure}

\Cref{fig:survey} shows the results of the survey.
We see that \emph{all} respondents deem 5ms gaps as acceptable, while 79\% of respondents deem gaps of 10ms as acceptable.

\end{document}